\documentclass[epj]{svjour}
 \newcommand\beq{\begin{equation}}
 \newcommand\eeq{\end{equation}}
\usepackage{graphics}
\begin{document}
\title{Propagation of Fast Partons in the Nuclear Medium}
\author{Mikkel B. Johnson
}                     
\institute{Los Alamos National Laboratory, Los Alamos, NM 87544}
\date{Received: date / Revised version: date}
%
\abstract{The color dipole approach has been applied in the target rest frame to 
address the issues of transverse momentum broadening and energy loss of a fast quark
propagating in the nuclear medium.  A recent application of the theory to the FermiLab
E772/E866 experimental data,  determining the rate of energy loss
of a quark propagating in the medium to be 2 to 3 GeV/fm,  will be reviewed.  
Calculations for the transverse
momentum distribution will be presented, and the results will be compared to the 
E866 data.  The theory will be shown to compare favorably to the data, and these 
results will be shown to suggest that the momentum broadening of a quark is about twice
the generally accepted size.  
} 
\maketitle
\section{Introduction}
\label{intro}
Nucleon-nucleus (or deuteron-nucleus) reactions at high energy are important experimentally because they constitute 
the conventional background for the less well known dynamics of relativistic nucleus-nucleus collisions, 
which have been identified as a means for producing the quark-gluon plasma in the labratory at facilities such as 
the Relativestic Heavy Ion Collider (RHIC) at BNL,  and the Large Hadron Collider (LHC) at CERN.  To develop sufficient 
confidence in theory to interpret the more 
complicated phenomena in nucleus-nucleus collisions, it is vital that the theory be tested against these 
simpler data first.  

Shadowing, quark energy loss, and transverse momentum broadening are three central issues of parton propagation 
in nucleus-nucleus 
collisions that may be quantitatively addressed in $p+A$ reactions, specifically in Drell-Yan (DY) reactions.  The 
color dipole approach in the target rest frame provides an attractive means for interpreting these data to learn
more about these phenomena.  After a brief description of the status of this theory, I will review 
recent results that are
improving our understanding of the physics behind these aspects of parton propagation in the nuclear medium.
  
Drell-Yan in the target rest frame may be described as a quark of the incident hadron 
interacting with a target nucleon, radiating a virtual photon $\gamma*$ of mass $M$ that 
subsequently decays into the 
observed Drell-Yan dilepton pair $\bar ll$.  In the target rest frame, the $\gamma*$ is a constituent 
of projectile fluctions, which are ``frozen" by 
time dilation for a length of time $t_c$ given by the uncertainty relation,
\beq
t_c=\frac{2E_q}{M_{q\bar ll}^2-m_q^2}
\label{coherence1}
\eeq
where $E_q$ and $m_q$ refer to the energy and mass of the projectile quark 
and $M_{q\bar ll}^2$ is the square of the effective
mass of the fluctuation,
\beq
M_{q\bar ll}^2=\frac{m_q^2}{\alpha}+\frac{M^2}{1-\alpha}+\frac{k_T^2}{\alpha(1-\alpha)},
\label{coherence2}
\eeq
where $\alpha$ is the fraction of the light-cone momentum of the incident quark carried by the lepton pair, 
and $k_T^2$ is the square of the transverse momentum of the lepton pair.  The fluctuation lifetime $t_c$ 
is called the coherence time, and the DY reaction occurs when the $\gamma*$ of the fluctuation is released 
by an interaction between one of
the constituents of the fluctuation and a target nucleon.  In the color dipole approach, this interaction is
mediated by the color dipole cross section, $\sigma_{\bar qq}$.

When the scattering takes place on a nucleus, multiple interactions with target nucleons can give rise to 
various 
medium effects.  Two limiting cases should be distinguished, the short coherence length limit (SCL) reached
when the coherence length $\ell_c \equiv c \langle t_c\rangle$ is much smaller than the interparticle spacing
$d$, $\ell_c \ll d$ (in a heavy nucleus, $d\approx 2 fm$), and the long coherence length limit (LCL) reached 
when $\ell_c \gg  R_A$, where $R_A$ is the nuclear radius.  
The coherence length determines the physics and also 
controls the way that medium effects are taken into account.  The theory simplifies these important limits.  
The intermediate case is generally more difficult to describe; however, if 
$0<\l_c\ll R_A$, the result can be obtained by interpolating between these limits using the square of the 
longitudinal form factor, $F^2_A(q_c)$, where $q_c = 1/l_c$ is the longitudinal momentum transferred 
in the reaction.  The Green function method~\cite{krt} was developed to handle the more difficult
situations.  Our results make use of all these cases.

One important medium effect is shadowing, and a basic consideration is the connection 
between shadowing and the coherence length.  In the SCL there is no shadowing because the duration of
the fluctuation is so short that it has no time to interact with the medium.  In the LCL, which
is applicable to 
reactions at the LHC and at RHIC under certain kinematic conditions, there is maximal shadowing. 

Medium effects also give rise to quark energy loss and to transverse momentum broadening.  These both 
arise in the SCL as the quark from the incident hadron undergoes 
various additional interactions with nucleons of the nucleus before it radiates the $\gamma*$.
For the LCL, the quark may undergo interactions that lead not only to shadowing but also to
additional momentum broadening.  Nuclear effects giving rise to momentum broadening and shadowing
are believed to arise predominantly from the same color dipole cross section $\sigma_{\bar qq}(r_T)$ 
that mediates the DY reaction on a nucleon.

The DY data relevant to shadowing, energy loss, and momentum broadening in which we are
interested here is the $p+A$ FermiLab data from the E772/E866 collaboration.  This data corresponds to the
range of coherent length where the shadowing is weak and the interpolation formula described above 
applies.

\section{Drell-Yan Reaction on a Nucleon in the Target Rest Frame}
\label{sec:1}
\begin{figure}[tbh]              
\includegraphics{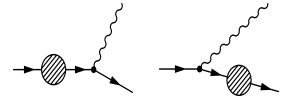}
\begin{center} 
\vspace{3cm}   
\parbox{8cm}                                       
{\caption[dyfig]                   
{\sl In the target rest frame, DY dilepton production looks like bremsstrahlung.  A quark or an anti-quark from the 
projectile hadron scatters off the target color field (denoted by the shaded circles) and
radiates a massive photon, which subsequently decays into the lepton pair.  The photon decay is not shown.  The 
photon can be radiated before or after the quark scatters.}
\label{dyfig}}
\end{center}                                                            
 \end{figure}

In the color dipole approach, the Drell-Yan  reaction occurs when a fast projectile quark scatters off the gluonic
field of the target, as shown in Fig.~\ref{dyfig}.  The color
dipole approach was originally proposed for deep-inelastic scattering (DIS).  Extension to Drell-Yan was developed by 
Kopeliovich~\cite{k} and subsequently by Brodsky, et al.~\cite{bhq}.  The DY reaction on a nucleon is described by 
\beq
M^2 \frac{d^2\sigma_{DY}}{dM^2dx_1}=\int_{x_1}^1dx_qF_q^h(x_q)\int 
d^2\rho |\Psi(\alpha ,\rho )|^2\sigma_{\bar qq}(\alpha \rho ), 
\label{sigdy}
\eeq
where $\sigma_{\bar qq}$ is the color-dipole cross section, $\rho$ is the transverse distance 
between the $\gamma*$ and
quark in the fluctuation,  and $\Psi$ represents the distribution for the incident quark to 
fluctuate into a quark and the  $\gamma*$.  In DY, the color dipole consists of the quark before and
after the release of the $\gamma*$, whose impact parameters differ by $\alpha\rho$. 
Here $x_q(x_1)$ is the fraction of light-cone momentum of the 
incoming hadron $h$ (for us a proton) carried by the quark (lepton pair); the light-cone momentum fraction $\alpha$
is given in terms of these variables, $\alpha =x_1/x_q$; and, $F_q^h$ is the quark distribution function 
of the incident hadron.

Deep-inelastic scattering from nucleons at HERA at DESY has been used to fix models of the color-dipole cross section.  
The HERA data suggests saturation, the
condition that the color dipole cross section approaches a constant value at large $r_T$,
\beq
\sigma_{\bar qq}(r_T)=\sigma_0(1-e^{-r_T^2/R_0^2}).
\label{cdm}
\eeq
Models that incorporate this saturation property include the GW~\cite{gbw} and KST~\cite{kst} models.  
We see that in both the GW and KST models, for small $r_T$, 
\beq
\sigma_{\bar qq}(r_T)\approx Cr_T^2,
\label{cdz}
\eeq
where $C=\sigma_0/R_0^2$ is a constant that depends on the quark energy, in the case of the KST model, and on Bjorken $x$ 
in the case of the GW model.  For us, the most important point is that once the color dipole 
cross section has been fit to experimental DIS data on the nucleon, the color-dipole formalism makes predictions for
Drell-Yan both in nucleon-nucleon~\cite{rpn} and nucleon-nucleus scattering with no further adjustment of parameters.  

The theory is 
expected to be valid for small Bjorken $x$ only, and specific calculations~\cite{rpn} have shown it to agree well 
for $x_2 < 0.1$ in both the magnitude and shape with the next-to-leading order parton model.  Although the calculations
disagreed with the E772 FermiLab data, they are in excellent agreement with 
the more recent E866 analysis in \cite{webb}.

\section{Medium Effects in DY Reactions for p+A Collisions in the Target Rest Frame}
\label{sec:2}

Medium effects are evident in nuclear ratios of cross sections $R^{A/A'}$.  These include analyses of both cross 
sections themselves, and the analysis of momentum distributions.  For each, there are two cases of interest:  weak 
shadowing and maximal shadowing.  

As we have stated, shadowing is controled by the 
coherence length, and explicit calculations in~\cite{mbj} show that for incident quarks
of the 800 GeV projectile protons the coherence length is quite a bit less than the nuclear radius, 
implying weak shadowing.  Additionally, 
these calculations show that the coherence length of the dominant $q-\gamma *$ fluctuation is nearly constant as
a function of $x_1$ and is therefore essentially a function of $x_2$ alone.  For $x_2 > .03$, $\ell_c< 2$ fm, 
and shadowing can be safely ignored.  For smaller $x_2$ the coherence length exceeds the interparticle spacing 
at the center of the nucleus, and shadowing therefore begins to become important.  The particular admixture of 
long- and short-coherence length contributions is determined by the longitudinal form factor, as stated in the 
introduction, and this depends in turn explicitly on the variation of $\ell_c$ with $x_1$ and $x_2$.

For cross section ratios, 
\beq
R^{A/A'}=\frac{d^2\sigma^A_{DY}}{dM^2dx_1}/\frac{d^2\sigma^{A'}_{DY}}{dM^2dx_1},
\label{R1}
\eeq
the E772/E866 FermiLab data at $E_p=800~GeV$ has been used for a recent 
determination of quark energy loss~\cite{mbj}.  The DY reaction at RHIC for $x_F=x_1-x_2>0.5$ and at the LHC
lies in the regime of long-coherence length physics, and such experiments at these facilities will be 
exploring a limit that has so far not been accessible experimentally.  Predictions of cross sections in $p+A$, $d+A$, 
and $A+A$ collisions~\cite{krtj} have been made recently in the latter case.

For momentum distribution ratios,
\beq
R^{A/A'}(p_T)=\frac{d^2\sigma^A_{DY}}{dp_T^2}/\frac{d^2\sigma^{A'}_{DY}}{dp_T^2},
\label{R2}
\eeq
data relevant to the the case of weak shadowing has been taken at $E_p=800~GeV$ by the FermiLab E772/E866 collaboration
and was used by them to study quark momentum broadening in nuclei, 
\beq
\delta\langle p_T^2\rangle =\langle p_T^2\rangle ^A- \langle p_T^2\rangle ^p.
\label{broad}
\eeq
This same regime is relevant to $p+A$ collisions at RHIC.  The case of maximal shadowing is relevant to the LHC and RHIC, 
and predictions are made in Ref.~\cite{krtj}.

\subsection{Cross Section Ratios}
\label{sec:2a}

Let us consider first the case of weak shadowing for the cross section ratios, as analyzed in \cite{mbj}.  
To calculate the cross section, we must calculate the cross section separately in the SCL and LCL and 
then interpolate between them using using the longitudinal form factor, as explained in the introduction.

The main medium effect in the short coherence length limit $x_2\gg 0.03$ the $\gamma *$ 
is the quark energy loss, which occurs before the $\gamma*$ is radiated.  
The interactions that cause the incident quark to lose energy reduce the quark momentum
fraction  $x_q$ and hence shift the value of $x_1=\alpha x_q$.  
The amount of reduction due to quark energy loss may be calculated as 
$x_q\approx E_q/E_h \rightarrow x_q-\kappa L/E_h$, where $L$ is the path length of the 
incident quark in the nucleus and $\kappa$ is its rate of energy loss.  Using these kinematic considerations, 
we find an expression in~\cite{mbj} for the DY cross section 
$M^2 \frac{d^2\sigma_{DY}^{(SLC)}}{dM^2dx_1}$ including its dependance
on $\kappa$.  It becomes sensitive to $\kappa$ for $x_1\approx 1$.

Mechanisms of energy loss~\cite{mbj}  consist of the primary ones of string breaking (SB), for which 
$\kappa_{SB} \approx 1~GeV/fm$ (numerically, the string tension), and the independent process of 
gluon radiation (GR), for which 
$\kappa_{GR}\approx 3\alpha_s\langle k_T^2\rangle 
\approx 0.8~GeV/fm$, where $\langle k^2_T\rangle\approx 0.65~GeV$ is the mean-square momentum of the radiated
gluon.  The induced energy-loss mechanisms arising from multiple interactions with nucleons are relatively
minor for nuclei of ordinary density and can be neglected.  We thus expect that $\kappa\approx 1.8~GeV/fm$.

As we have said, shadowing arises in the LCL through multiple interactions between the quark
and the nucleus following the emission of the $\gamma*$.  For weak shadowing, Glauber theory to second 
order in the number of interactions may be used, giving 
\beq
M^2 \frac{d^2\sigma_{DY}^{(LCL)}}{dM^2dx_1}=\langle 
\sigma_{\bar qq}(\alpha \rho )\rangle (1-\frac{\sigma_{eff}\langle T_A\rangle }{4})
\label{Glauber}
\eeq
where the brackets around the
color dipole cross section refer to the averages over $(\rho,\alpha)$ in Eq.~(\ref{sigdy}), and where
$\langle T_A\rangle$ is the average of the thickness function $T_A(b)$ over impact parameter $b$, 
\beq
<T_A>=\frac{1}{A}\int d^2bT_A^2(b),
\label{avTA}
\eeq
with $T_A(b)=T_A(b,\infty)$, where
\beq
T_A(b,z)=\int_{-\infty}^z d^2b\rho_A(b,z),
\label{TA}
\eeq
and $\rho_A(b,z)$ is the nuclear density.  The quantity 
$\sigma_{eff}=\langle\sigma_{\bar qq}^2\rangle/\langle\sigma_{\bar qq}\rangle$ in Eq.~(\ref{Glauber})
is obtained by expanding the multiple scattering through second order in $\sigma_{\bar qq}$. In the 
color-dipole approach shadowing is completely determined by the color dipole cross section when
shadowing is weak. 

Interpolating between the LCL where $x_2\ll 0.03$ and the SCL where $x_2\gg 0.03$, we obtain
\beq
 \frac{d^2\sigma_{DY}}{dM^2dx_1}= \frac{d^2\sigma_{DY}^{(SCL)}}{dM^2dx_1}(1-F^2_A(q_c))
+ \frac{d^2\sigma_{DY}^{(LCL)}}{dM^2dx_1}F^2_A(q_c),
\label{interpolate}
\eeq
where $q_c=1/\ell_c$ and the longitudinal form factor is defined as
\beq
F_A^2(q)=\frac{1}{<T_A>}\int d^2b|\int^\infty_{-\infty} dze^{iqz}\rho(b,z)|^2.
\label{longff}
\eeq
Note that this interpolation formula has the correct limits for $\ell_c\rightarrow 0$ and $\ell_c\rightarrow\infty$.  
It also correctly reproduces the Gribov shadowing formula~\cite{gribov,kk} and correctly describes shadowing 
in DIS~\cite{mbj}.  Because the cross section becomes sensitive to both shadowing and $\kappa$ for $x_1\approx 1$,
it is important that both effects are included in the theory (note that the depencence of shadowing on $\ell_c$ 
becomes quite intricate in this limit~\cite{krt2,mbj}).  This theory then provides a quantitative test of 
the value of the relatively poorly known quark energy loss in the nuclear medium, since shadowing is completely 
determined by theory.  

\begin{figure}[tbh]              
\includegraphics{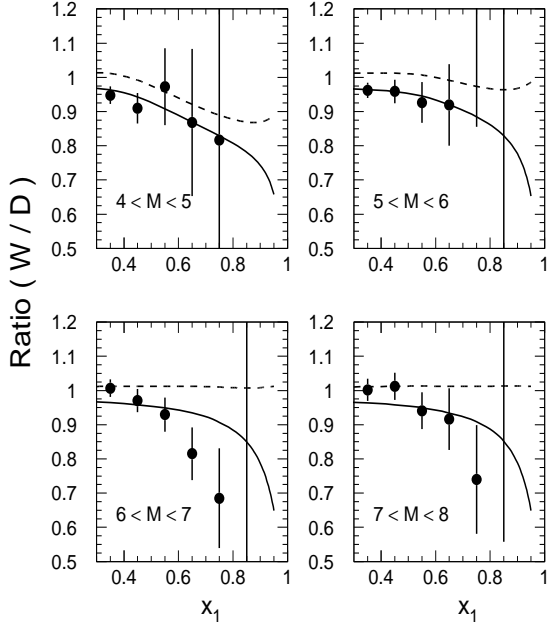}
\begin{center} 
\vspace{8cm}   
\parbox{8cm}                                       
{\caption[shad1]                   
{\sl Examples of ratios of the DY cross sections of tungsten to deuterium as functions of $x_1$ for various 
intervals of $M$.  Dashed curves correspond to net shadowing contribution, solid curves show the 
full effect including shadowing and energy loss.  Data are from Refs.~\cite{alde,unp}.}
\label{w-d}}
\end{center}                                                            
 \end{figure}

For comparing the theory to experiment, we have evaluated Eq.(\ref{interpolate}) and compared to the E866/E772 
FermiLab data.  This data provides the dependence of $R^{A/A'}$ over a range of $(A,x_1,M^2)$; our ability to
reproduce these data gives
us confidence that we are able to separate the effects of shadowing and energy loss.  Typical results are
shown in Fig.~\ref{w-d}.

The rate of energy loss corresponding to the fit to the the data, as shown in Fig.~\ref{w-d}, 
is 
\beq
\kappa=2.73\pm 0.37 GeV/fm
\label{eloss}
\eeq
Note that the value in Eq.~(\ref{eloss}) is somewhat larger than the theoretical estimate 
of $1.8~GeV/fm$ given above; it is also considerably larger than values found in previous analyses 
of the DY data~\cite{vasiliev}.  We regard our value as more reliable, since
the color dipole approach is able to disentangle shadowing from energy loss using a reliable theoretical 
calculation of the shadowing contribution.  
\subsection{Momentum Distribution Ratios~\cite{jk}}
\label{sec:2b}
Calculations show that the momentum distributions from E772/E866 experiments with $800 GeV$ protons or 
measurements at RHIC with $s^{1/2}=200~GeV$ protons at $x_F=0$ correspond to the short coherence length 
limit, as long as $x_2>0.05$.  Working in the short coherence length limit simplifies the 
theory by eliminating the shadowing term in Eq.~(\ref{interpolate}).  
We will restrict our attention to this kinematic regime, which is possible because the data is 
available binned in $x_2$ intervals, and thus concentrate on the first (LCL) term in this expression.

It was recognized in Ref.~\cite{jkt}
that at high energies, the multiple interaction of a quark propagating in the nuclear medium can be eikonalized and
exponentiates into a factor containing the color dipole cross section and nuclear thickness function.  The physics 
of momentum broadening of the quark is therefore that of color filtering, or absorption of large dipoles leading
to diminished transverse separation with distance.  Then, the probability distribution $W^A(k_T)$ that a quark 
will have acquired transverse momentum $k_T$ at a position $(b,z)$ in the nucleus $A$ becomes~\cite{jkt}
\beq
W^A(k_T)=\frac{1}{(2\pi )^2}\int d^2r_Te^{ik_T\cdot r_T}e^{-\frac{1}{2}\sigma_{\bar qq}(r_T)T_A(b,z)},
\label{p-distr}
\eeq  
where $T_A(b,z)$ is defined in Eq.~(\ref{TA}).
We express the cross section, $\sigma_{DY}^A(p_T)$, for a proton to produce a DY pair on a nucleus $A$ with 
transverse momentum $p_T$ as  convolution of the probability $W^A(k_T)$ and the
DY momentum distribution on a proton.  The resulting expression entails an integral over $\Psi(\alpha,\rho)$ 
as in Eq.~({\ref{sigdy}), and for the results we show below, this integral has been
performed numerically.  However, for the purpose of discussing the result here, we replace the quantities that
depend on $\alpha$ by averages.  Then, the ratio $R^{A/p}(p_T)$ may be expressed as in terms of the
DY cross section $\sigma_{DY}(p_T)$,
\beq
R^{A/p}(p_T)=\frac{1}{\sigma_{DY}^p(p_T)}\int d^2k_T W^A(k_T)\sigma_{DY}^p(p_T-\bar \alpha k_T)
\label{convolution}
\eeq
where $\bar \alpha$ arises from the fact that the DY pair carries away a fraction $\alpha$ of the
transverse momentum of a quark in the incident proton.  We find that 
$\bar \alpha = \langle \alpha ^2\rangle ^{1/2}\approx 0.97$.  The average over $\alpha$
implicit in Eq.~(\ref{convolution}) also entails an evaluation of the quark energy, $\langle E_q\rangle\approx E_q/3$,
since we use the GW model of $\sigma_{\bar qq}$, which depends explicitly on this energy.   

For our calculations below we have chosen 
\beq
\sigma_{DY}^p(p_T)\propto (1+p_T^2/\Lambda^2)^{-6}
\label{sig-p}
\eeq
where the value of $\Lambda^2= 7 GeV^2$ is taken from Ref.~\cite{pat}.  

Some of the integrals in Eqs.(\ref{p-distr},\ref{convolution}) may be done analytically by 
expanding out the exponential in Eq.~(\ref{cdm}).  We find that to an excellent approximation, the
effect of averaging Eq.~(\ref{p-distr}) over $(b,z)$ is to replace 
\beq
T_A(b,z)\rightarrow \langle T_A\rangle /2.  
\label{avT}
\eeq
When evaluating Eq.~(\ref{p-distr}), we also
take into account gluon shadowing by replacing 
$\sigma_{\bar qq}\rightarrow R_G\sigma_{\bar qq}$, where
$R_G$ is the gluon shadowing function.  See Ref.~\cite{jkt} for more discussion.  Our calculations
are however insensitive to $R_G$.
   
Our calculation for $R^{W/Be}(p_T)$ is shown in Fig.~\ref{tmd1}.  The data relevant 
to $p_T$ distributions on nuclei are FermiLab E772/E866 data~\cite{vasiliev,unp,leitch}.  
 \begin{figure}[tbh]              
\includegraphics{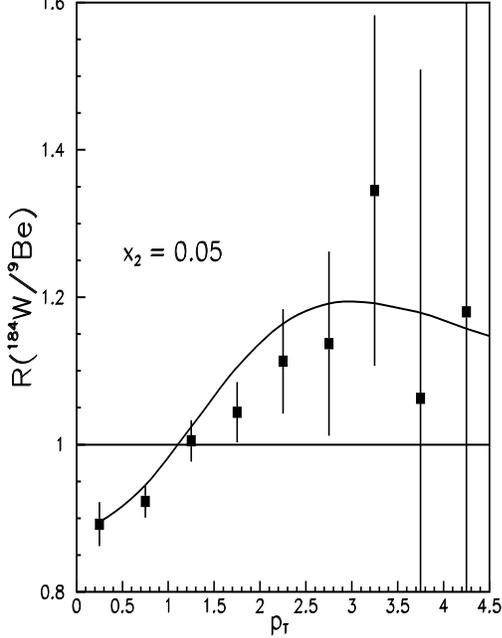}
\begin{center} 
\vspace{8cm}   
\parbox{8cm}                                       
{\caption[shad1]                   
{\sl Comparison of theoretical prediction of $R^{W/Be}(p_T)$ vs. $p_T$ (in $GeV/c$) to experiment 
for $x_2 = 0.05$.  Data are from the FermiLab E772/E866 collaboration~\cite{vasiliev,unp,leitch}.}  
\label{tmd1}}
\end{center}                                                            
 \end{figure}  
One sees that the calculation is in quite good agreement with experiment.  Note that the details of
the Cronin effect, the rise above $1$ for $p_T\approx 3$ in Fig.~\ref{tmd1}, are completely explained 
with no adjustable parameters.  We find comparable 
agreement with $R^{Fe/Be}(p_T)$.
\subsection{Transverse Momentum Broadening}
\label{sec:2c}
The transverse momentum broadening of a quark propagating in a nucleus is defined in Eq.~(\ref{broad})
where the mean momentum is defined as
\beq
\langle p_T^2\rangle =\frac{\int d_{pT}^2 p_T^2 \sigma_{DY}^A(p_T)}{\int d_{pT}^2 \sigma_{DY}^A(p_T)}.  
\label{meanp}
\eeq
The dependence of the quantity $\delta\langle p_T^2\rangle$ on $A$ has been studied experimentally using the E772 data 
in Ref.~\cite{mmp}.  They found that for large A, 
\beq
\delta\langle p_T^2\rangle^A_{exp}\approx 0.021 A^{1/3} GeV^2.
\label{exp}
\eeq

We can make a calculation of the same quantity in the theory just described.  From Eq.~(\ref{convolution}) we easily
find that 
\beq
\sigma_{DY}^A(p_T)=\frac{1}{\bar\alpha^2}\int d^2r_Te^{ip_T\cdot r}W^A(r_T)\sigma_{DY}^p(r_T)
\label{sig1}
\eeq
where
\beq
W^A(r_T)=e^{-\frac{1}{2}\sigma_{\bar qq}(r_T)T_A(b,z)}
\label{mom1}
\eeq
and 
\beq
\sigma_{DY}^A(r_T)=\int d^2p_Te^{-p_T\cdot r_T}\sigma_{DY}^A(p_T)
\label{mom2}
\eeq
characterizes how a quark acquires transverse momentum as it propagates through the 
nucleus.  From Eq.~(\ref{meanp},\ref{sig1}) we find
\beq
\langle p_T^2\rangle^A=-\frac{\nabla^2\sigma_{DY}^p(r_T)}{\sigma_{DY}^p(r_T)}|_{r_T=0}-
\frac{\nabla^2W^A(\bar\alpha r_T)}{W^A(\bar\alpha r_T)}|_{r_T=0}
\label{ptA}
\eeq
The first term in Eq.~(\ref{ptA}), the mean-square momentum on a nucleon, is formally infinite.
Fortunately, the nucleon contribution gets subtracted to obtain $\delta\langle p_T^2\rangle$,  
\beq
\delta\langle p_T^2\rangle =-\bar\alpha^2\frac{\nabla^2W^A(r_T)}{W^A(r_T)}|_{r_T=0}.
\label{delpt1}
\eeq
Note that the shadowing contribution in Eq.~(\ref{interpolate}), which we justified in dropping
by going to $x_2>0.05$, leads to a divergent contribution that no longer cancels exactly;
in cases where FSI are important other means are needed to define a meaningful quantity~\cite{krtj}.  

Using Eq.~(\ref{mom1}), along with the observation that $\sigma_{\bar qq}(r_T)$ is proportional
to $r_T^2$ at small $r_T$ (see Eq.~(\ref{cdz})),
\beq
\delta\langle p_T^2\rangle=\bar\alpha^2C\langle T_A\rangle
\label{final}
\eeq
where we have used the average of $T_A$ as given in Eq.~(\ref{avT}). 

We may now make estimates to compare with the experimental result in Eq.~(\ref{exp}).  Using values 
of $C\approx 4-5$ as obtained from the GW or KST models, and taking the sharp-surface model for
the nuclear density,
\beq
\langle T_A\rangle = \frac{3}{2}\rho_0R_A
\label{sharp}
\eeq
where $\rho_0\approx 0.16 fm^{-3}$ is the central density of heavy nuclei and 
$R_A\approx 1.1 A^{\frac{1}{3}}$, we find from theory that
\beq
\delta\langle p_T^2\rangle^A_{Th}\approx 0.047A^{\frac{1}{3}}~GeV^2
\label{th}
\eeq
The theoretical result in Eq~(\ref{th}) is clearly larger than and in 
apparent disagreement with the
experimental result in Eq.~(\ref{exp})~\cite{jkt,r}.  However, it has been pointed out~\cite{jm} 
that large systematic errors must be included in Eq.~(\ref{exp})
and that when these are taken into account,
the result in Eq.~(\ref{th}) could be accommodated.

\subsection{Predictions for Maximal Shadowing}
\label{sec:3c}
As we have stated, the case of maximal shadowing corresponds to $\ell_c\gg R_A$, and experiments
in this regime have not yet been possible.  The color dipole approach in the target rest frame
applies in this limit, and numerous predictions have been presented in Ref.~\cite{krtj}.  We give
one example in Fig.~\ref{longc}, namely the DY ratios as a function of transverse momentum presented separately
for longitudinal (L) and transverse + longitudinal ($T+L$) photon polarization.

\begin{figure}[tbh]              
\includegraphics{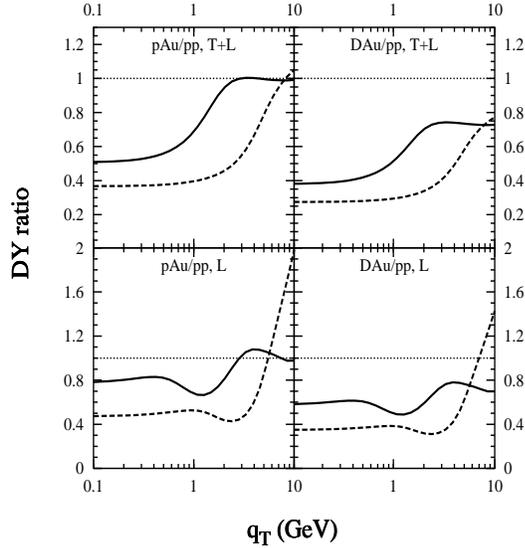}
\begin{center} 
\vspace{8cm}   
\parbox{8cm}                                       
{\caption[long]                   
{\sl Nuclear effects on the DY transverse momentum distribution.
  Curves show the DY cross sections for $pAu$ (left) and deuterium --
gold (right) collisions per nucleon divided                                    
  by  the DY cross section from $pp$ scattering.                             
Solid curves are predictions for RHIC ($\sqrt{s}=200$ GeV)  and    
dashed for LHC ($\sqrt{s}=5.5$ TeV). Calculations are for $M$=4.5 GeV and $x_F=$ 0.5.}
\label{longc}}
\end{center}                                                            
 \end{figure}  

In contrast to the weak shadowing limit discussed earlier, transverse momentum broadening here 
is associated with interactions with the quark following the emission of $\gamma*$.  
Note that the Cronin peak is much less visible in the LCL,
which is a consequence of the gluon shadowing.  For evaluating the gluon shadowing here,
the Green's function method is required, and we take $R_G$ from the calculations of Ref.~\cite{kst2}.

\section{Summary and Conclusions}
\label{conc}
We have shown that the color dipole approach formulated in the target rest frame has 
advantages for examining nuclear modifications in $p+A$ collisions because the effects of 
shadowing may
be determined theoretically.  In the SCL, $\ell_c\ll R_A$, energy loss and momentum broadening 
are mediated by multiple interactions of the incident quark with target nucleons before the 
emission of the $\gamma*$.  Analysis
of the E772/E866 FermiLab experimental DY data at $E_p=800~GeV$ in the color dipole 
approach has provided improved determinations of these quantities, giving
\beq
\frac{dE_q}{dz}=-2.73\pm 0.35~GeV/fm
\label{elossval}
\eeq
and
\beq
\delta\langle p_T^2 \rangle\approx 0.047A^{\frac{1}{3}}~GeV^2
\label{ptval}
\eeq
The value of the rate of energy loss is larger than previous determinations, and the
value of $\delta\langle p_T^2\rangle$ is about twice as large as the conventional
value.  

Predictions of the transverse momentum dependence of DY cross sections, for $p+A$ and $A+A$ 
collisions, have been made in the color dipole approach.  These provide tests of the theory
in a new physical regime of strong shadowing, where $\ell_c\gg R_A$, that may be explored 
at the LHC in the future.  
\section{Acknowledgements}
I would like to acknowledge the essential contributions of my theoretical and experimental
collaborators B. Kopeliovich, A. Tarasov, J. Raufeisen J. Moss, M. Leitch, 
and P. McGaughey.  This work was supported in part by U.S. Department of Energy.


\begin{thebibliography}{}
%
%
\bibitem{krt} B. Z. Kopeliovich J. Raufeisen, and A. V. Tarasov, Phys. Lett. B \textbf{440}, 
(1998) 151; J. Raufeisen, A. V. Tarasov, and O. O. Voskresenskaya, Eur. Phys. J. \textbf{A5} (1999) 173.
\bibitem{k} B. Z. Kopeliovich, Proc. of the Workshop Hirschegg '95:  Dynamical Properties of Hadrons in Nuclear 
Matter, Hirschegg, January 16-21, 1995, ed. by H. Feldmeyer and W. N\"{o}renberg, Darmstadt, 1995, p. 102 (hep-ph/9609385).
\bibitem{bhq} S. J. Brodsky, A. Hebecker, and E. Quack, Phys. Rev. D\textbf{55} (1997) 2584.
\bibitem{gbw} K. Goltec-Biernat and M. W\"ustoff, Phys. Rev. D\textbf{59} (1999) 014017 (hep-ph/9807513);
Phys. Rev. D\textbf{60} (1999) 114023 (hep-ph/9903358).
\bibitem{kst}B.Z.~Kopeliovich, A.~Sch\"afer and A. V.~Tarasov, Phys. Rev. C{\bf 59} (1999) 1609 (hep-ph/9908245).
\bibitem{rpn} J. Raufeisen, J-C. Peng, and G. Nayak, Phys. Rev. D\textbf{66} (2002), 034024 (hep-ph/0204095); B. Z.
Kopeliovich, J. Raufeisen, and A. V. Tarasov, Phys. Lett. \textbf{B503} (2001) 91; M. A. Betemps, M. B. Gay Ducati, 
M. V. T Machado, and J. Raufeisen, Phys. Rev.D \textbf{67} (2003) 114008.
\bibitem{webb} J. Webb, New Mexico State University Thesis (2002) (hep-ex/0301031).
\bibitem{mbj} M. B. Johnson, et al., Phys. Rev. Lett. \textbf{86} (2001) 4483 (hep-ex/0010051); and M. B. Johnson, et al., 
Phys. Rev. C\textbf{65} (2002) 025203 (hep-ph/0105195).
\bibitem{krtj} B. Z. Kopeliovich, J. Raufeisen, A. V. Tarasov, and M. B. Johnson,
Phys. Rev. C\textbf{67} (2003) 025203 (hep-ph/0105195).
\bibitem{gribov} V. N.~Gribov, Sov. Phys. JETP {\bf 29} (1969) 483;                                    
{\bf 30} (1970) 709
\bibitem{kk} V.~Karmanov and L. A.~Kondratyuk, Sov. Phys. JETP
Lett. {\bf 18} (1973) 266.
\bibitem{krt2} B. Z. Kopeliovich, J. Raufeisen, and A. V. Tarasov, Phys. Rev. C\textbf{62} (2000) 035204.
\bibitem{alde} D. M. Alde et al., Phys. Rev. Lett. {\bf 66} (1991) 133. 
\bibitem{unp} Unpublished E772 data.
\bibitem{vasiliev} M. A. Vasiliev, et al., Phys. Rev. Lett. \textbf{83} (1999) 2304.
\bibitem{jk}  Results shown in this subsection are preliminary.  M. B. Johnson and B. Z. Kopeliovich, unpublished.
\bibitem{jkt} M. B. Johnson, B. Z. Kopeliovich, and A. V. Tarasov, Phys. Rev. C\textbf{63} (2001), 035203 (hep-ph/0006326).
\bibitem{pat} Erratum to P. McGaughey, et al., Phys. Rev. D\textbf{50} (1994) 3038. 
\bibitem{leitch} M. Leitch, private communication. 
\bibitem{mmp} P. L. McGaughey, J. M. Moss, and J- C. Peng, Annu. rev. Nucl. Part. Sci. \textbf{49} (1999) 217.
\bibitem{r} J. Raufeisen, Phys. Lett. \textbf{B557} (2003) 184.
\bibitem{jm} Joel Moss, private communication.
\bibitem{kst2} B. Z. Kopeliovich, A. Sch\"afer, and A. V. Tarasov, Phys. Rev. D \textbf{62} (2000) 054022.
(hep-ph/9908245).
\end{thebibliography}
\end{document}